\documentclass[aps, prl, reprint, superscriptaddress,preprintnumbers,amssymb]{revtex4-2}
\usepackage{graphicx}
\usepackage{dcolumn}
\usepackage{bm}

\begin{document}

\preprint{ver1.2}

\title{Fermi surface and field-induced Lifshitz transition in the antiferromagnet NbMnP exhibiting an anomalous Hall effect}

\author{Taichi Terashima}
\email{TERASHIMA.Taichi@nims.go.jp}
\affiliation{Research Center for Materials Nanoarchitectonics (MANA), National Institute for Materials Science, Tsukuba 305-0003, Japan}
\author{Yuya Hattori}
\altaffiliation[Present address: ]{Department of Physics, Tokyo Metropolitan University, 1-1, Minami-osawa, Hachioji 192-0397, Japan}
\affiliation{Center for Basic Research on Materials (CBRM), National Institute for Materials Science, Tsukuba 305-0047, Japan}
\author{David Graf}
\affiliation{National High Magnetic Field Laboratory, Florida State University, Tallahassee, Florida 32310, USA}
\author{Yuki Arai}
\author{Hitoshi Sugawara}
\author{Hisashi Kotegawa}
\email{kotegawa@crystal.kobe-u.ac.jp}
\affiliation{Department of Physics, Kobe University, Kobe, Hyogo 657-8501, Japan}
\author{Ibuki Terada}
\affiliation{Department of Materials Science, Graduate School of Engineering, Osaka Metropolitan University, Sakai, Osaka 599-8531, Japan}
\author{Michi-To Suzuki}
\email{mts@omu.ac.jp}
\affiliation{Department of Materials Science, Graduate School of Engineering, Osaka Metropolitan University, Sakai, Osaka 599-8531, Japan}
\affiliation{Center for Spintronics Research Network, Graduate School of Engineering Science, The University of Osaka, Toyonaka, Osaka 560-8531, Japan}

\date{\today}

\begin{abstract}
We performed Shubnikov--de Haas (SdH) oscillation measurements on the $\mathbf{Q}=0$ antiferromagnet NbMnP, which exhibits a large anomalous Hall effect.
The SdH frequencies observed in the low-field region up to 17.5~T are well described by first-principles band-structure calculations based on the $B_{3g}+B_{2u}$ noncollinear magnetic structure determined by neutron diffraction.
The effective masses are enhanced by approximately a factor of two relative to the calculated band masses.
High-field measurements up to 41.5~T reveal a Lifshitz transition accompanied by a pronounced evolution of the SdH frequency spectrum.
These results confirm that first-principles calculations provide a reliable tool for the development of functional antiferromagnetic materials, while also demonstrating that field-induced changes in the Fermi surface itself need to be taken into account in understanding their high-field magnetotransport properties.
\end{abstract}

\maketitle

Materials that are basically antiferromagnetic yet possess spin-split electronic band structures have attracted considerable attention.
One motivation is the possibility that such materials may exhibit outstanding functionalities in applications such as spintronics.
Notable examples include the reports of spontaneous anomalous Hall effects in Mn$_3$Sn \cite{Nakatsuji15Nature} and hexagonal FeS \cite{Takagi25NatMat}.
The former is a noncollinear antiferromagnet, whereas the latter is a collinear antiferromagnet, neither of which would be expected to exhibit a spontaneous anomalous Hall effect within the conventional picture.
In this rapidly developing field, electronic band structures obtained from first-principles calculations and the Berry curvature calculated from them have become important guides for materials development.
In the present work, we measured quantum oscillations in high-quality single crystals of the antiferromagnet NbMnP, which exhibits a spontaneous anomalous Hall effect, and tested the first-principles band-structure calculations against the experimental results.

NbMnP crystallizes in an orthorhombic structure with space group Pnma and undergoes an antiferromagnetic transition at $T_{\mathrm N}=234$~K \cite{Matsuda21PRB, Zhao22Magnetism}.
According to neutron diffraction measurements \cite{Matsuda21PRB}, its magnetic structure is noncollinear with a propagation vector $\mathbf{Q}=0$, meaning that the magnetic unit cell coincides with the crystallographic unit cell, and is described by a superposition of the two irreducible representations $B_{3g}$ and $B_{2u}$, with an ordered magnetic moment of 1.2~$\mu_{\mathrm B}$ per Mn (see the Supplemental Material for the magnetic structure \cite{*[][ for the magnetic structure{,} band-structure calculations{,} SdH oscillations in sample R{,} and effective mass determination{,} which includes Refs{.} 26--31.] SM})).
Canting of the magnetic moments gives rise to a very small ferromagnetic component of approximately $10^{-3}~\mu_{\mathrm B}$/Mn.
Kotegawa \textit{et al.} \cite{Kotegawa23npjQM} observed a large spontaneous anomalous Hall effect in NbMnP, with $\sigma_{yz}\sim230~\Omega^{-1},\mathrm{cm}^{-1}$, comparable to that observed in Mn$_3$Sn \cite{Nakatsuji15Nature} or Mn$_3$Ge \cite{Nayak16SciAdv}.
The residual resistivity ratio (RRR) of the single crystals initially grown by a self-flux method was approximately 2.
Subsequently, a crystal-growth method using Ga flux was developed, yielding high-quality single crystals with RRRs as high as 42 \cite{Arai24JPSJ}.
This improvement enabled a systematic comparison of the anomalous Hall conductivity among samples with different RRRs.
The anomalous Hall conductivity was found to remain nearly unchanged at low temperatures, in the range of 230–450~$\Omega^{-1}\mathrm{cm}^{-1}$, despite variations in the conductivity, indicating that the anomalous Hall effect in NbMnP originates from an intrinsic mechanism.
Band-structure calculations for the $B_{3g}+B_{2u}$ magnetic structure yielded Berry-curvature-derived anomalous Hall conductivities of 273 or 366~$\Omega^{-1}\mathrm{cm}^{-1}$, in reasonable agreement with the experimental values \cite{Kotegawa23npjQM, Terada26PRM}.

In this work, we report Shubnikov–de Haas (SdH) oscillations observed in the magnetoresistance of NbMnP single crystals grown by the Ga-flux method.
The SdH frequencies observed in the low-field region are well described by band-structure calculations based on the $B_{3g}+B_{2u}$ magnetic structure.
At high fields, however, the SdH frequencies evolve with magnetic field, revealing a change in the topology of the Fermi surface, namely, a Lifshitz transition.

High-quality single crystals of NbMnP were grown by a Ga-flux method as detailed in Ref.~\cite{Arai24JPSJ}.
The resulting crystals were needle-shaped and elongated along the $b$ axis.
For four-probe electrical resistivity measurements, gold wires were first attached to the sample using silver paste, followed by pulsed-current spot welding to reduce the contact resistance.
The measurement current was applied along the $b$ axis.
Magnetoresistance measurements for observing SdH oscillations were carried out between 7 and 17.5~T using a dilution refrigerator at NIMS and between 15 and 41.5~T using a $^3$He refrigerator at NHMFL.
Because of limited magnet time, measurements at lower magnetic fields were performed for only a few field orientations at both NIMS and NHMFL.
Two samples, denoted as B and R, were measured at both NIMS and NHMFL.
The residual resistivity ratios (RRRs) of samples B and R were 65 and 61, respectively.
In addition, several other samples were measured at NIMS, and the observed SdH oscillations were found to be consistent among all samples.

The electronic structure calculations were performed using the QUANTUM ESPRESSO package \cite{GiannozziJPCM09} (see the Supplemental Material for details) \cite{Kotegawa23npjQM, Terada26PRM}. 
Exchange correlation energy was treated within the generalized gradient approximation (GGA) parametrized by Perdew, Burke, and Ernzerhof \cite{Perdew96PRL}.
Spin-orbit coupling was included in the calculations.
The magnetic structure was constrained to the $B_{3g}+B_{2u}$ noncollinear configuration determined by neutron diffraction.

\begin{figure*}
\includegraphics[width=16cm]{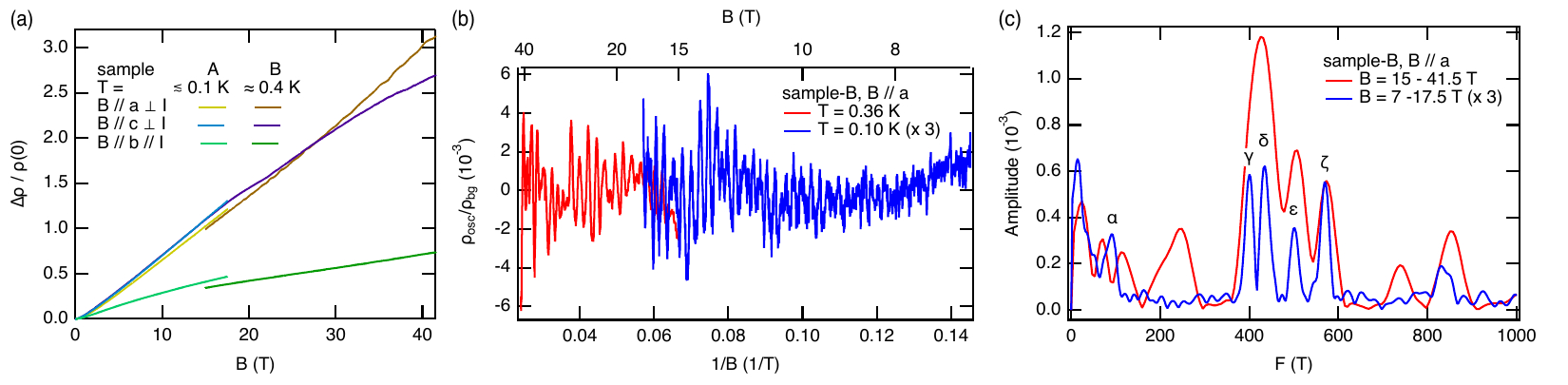}
\caption{\label{Fig1} 
(a) Magnetoresistance of NbMnP, defined as $\Delta\rho/\rho(0)=[\rho(B)-\rho(0)]/\rho(0)$.
Data for sample A up to 17.5~T and for sample B from 15 to 41.5~T are shown.
(b) SdH oscillations $\rho_{osc}$ obtained by subtracting a smooth background $\rho_{bg}$ from the magnetoresistance for $B \parallel a$ of sample B.
The data measured at NIMS in the field range from 7 to 17.5~T and at NHMFL from 15 to 41.5~T are shown.
(c) Fourier transform spectra of the SdH oscillations shown in (b).
Because the inverse-field window, $\Delta(1/B)$, is larger for the NIMS data than for the NHMFL data, the former provides a higher frequency resolution.
The fundamental frequency resolutions, $(\Delta(1/B))^{-1}$, are 12~T for the NIMS data and 23~T for the NHMFL data.
The major peaks in the low-field NIMS spectrum are labeled with Greek letters.
}
\end{figure*}

Figure~1(a) shows the magnetoresistance of sample B for current applied along the $b$ axis in magnetic fields up to 41.5~T for the field orientations $B \parallel a$, $B \parallel c$, and $B \parallel b$.
Since the NHMFL data are available only above 15~T, they are shown together with the NIMS data of sample A (RRR = 72) measured up to 17.5~T.
The relatively small magnetoresistance for $B \parallel b$ arises from the longitudinal magnetoresistance configuration.
For the transverse magnetoresistance, a tendency toward saturation is observed for $B \parallel c$, whereas for $B \parallel a$ the magnetoresistance increases nearly linearly up to 41.5~T.
This nearly linear transverse magnetoresistance is reminiscent of similar behavior frequently observed in topological semimetals \cite{Liang15nmat, Novak15PRB, Singha17PNAS}.
However, the absolute magnitude of the magnetoresistance in the present case is much smaller than that typically reported for such materials.

Figure~1(b) shows the SdH oscillations obtained by subtracting a smooth background from the magnetoresistance of sample B for $B \parallel a$.
The data measured at NIMS in the field range from 7 to 17.5~T and at NHMFL from 15 to 41.5~T are shown.
Figure~1(c) shows the corresponding Fourier transform spectra.
Multiple frequency peaks are observed.
The low-field NIMS data and the high-field NHMFL data are in good agreement in the frequency range of $F=400$--600~T.
However, additional frequency peaks emerge in the high-field data.

\begin{figure}
\includegraphics[width=6.5cm]{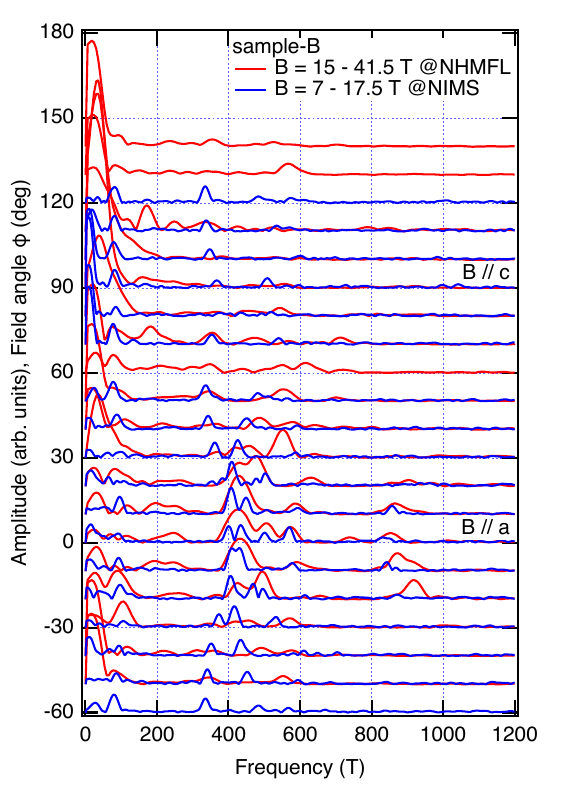}
\caption{\label{sample_B} 
Fourier transform spectra of the SdH oscillations in NbMnP sample B for various magnetic-field orientations within the $ac$ plane.
Both the low-field NIMS data and the high-field NHMFL data are shown.
The field angle $\phi$ is measured from the $a$ axis, and successive spectra are vertically offset according to $\phi$ for clarity.
}
\end{figure}

\begin{figure*}
\includegraphics[width=18cm]{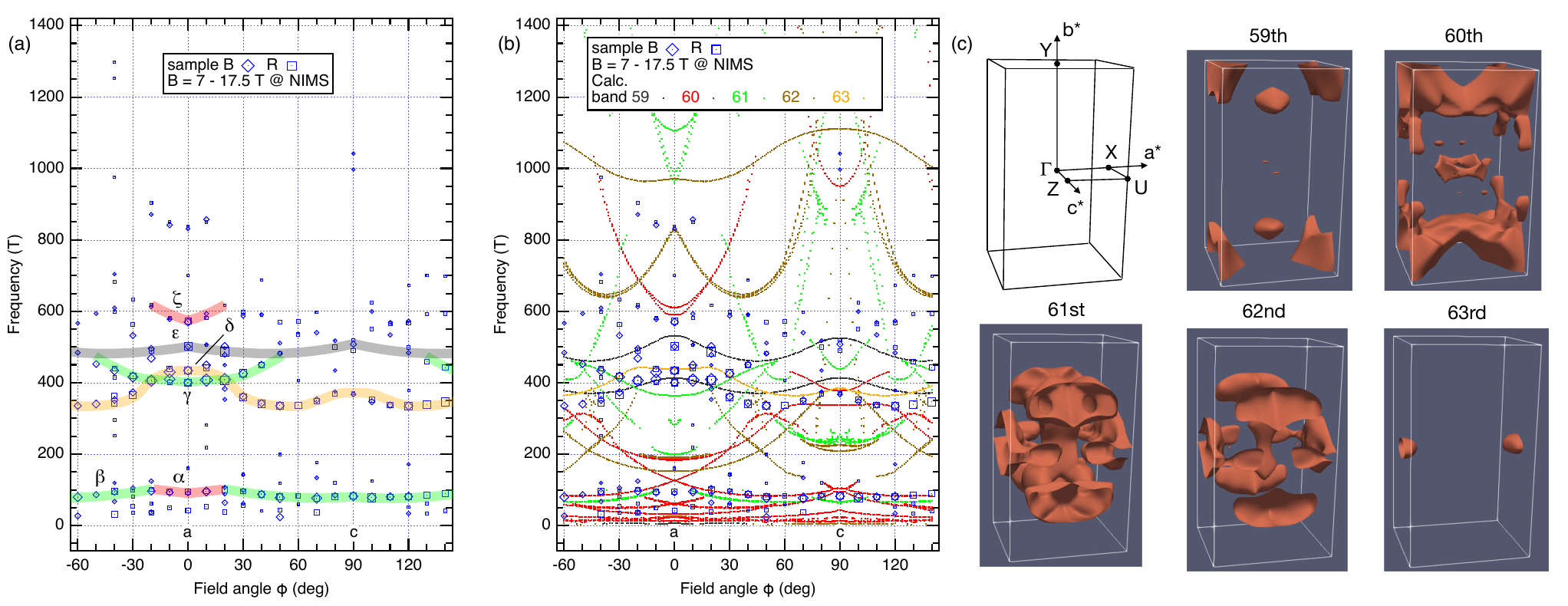}
\caption{\label{sample_B} 
(a) Angular dependence of the SdH frequencies in NbMnP.
The results for samples B and R are shown together.
The Fourier transforms were performed over the field range from 7 to 17.5~T.
The marker size reflects the oscillation amplitude.
The frequency branches $\alpha$, $\beta$, $\gamma$, $\delta$, $\varepsilon$, and $\zeta$ are indicated.
(b) Comparison of the experimental data shown in (a) with the frequencies predicted by the band-structure calculations.
(c) Fermi surfaces obtained from the band-structure calculations.
}
\end{figure*}

Figure~2 shows the Fourier transform spectra measured at low and high magnetic fields for various field orientations within the $ac$ plane.
The differences between the low-field and high-field spectra are particularly pronounced around $\phi = 30^\circ$, where $\phi$ is measured from the $a$ axis.
The field dependence of the Fourier spectra suggests that the Fermi surface changes with magnetic field, a point that will be revisited later.
Therefore, the low-field data are considered more appropriate for comparison with the zero-field Fermi surface obtained from band-structure calculations.

Accordingly, the frequencies extracted from the low-field data are plotted as a function of field angle in Fig.~3(a).
To demonstrate reproducibility, the results for sample R are shown together with those for sample B.
(Details of the SdH oscillations in sample R are provided in the Supplemental Material.)
Figures 2 and 3(a) show that the angular dependence of the SdH oscillations is symmetric about both $B \parallel a$ and $B \parallel c$ as expected from the orthorhombic crystal symmetry.

Figure~3(c) shows the Fermi surfaces obtained from the band-structure calculations.
Because the $B_{3g}+B_{2u}$ magnetic order breaks the combined $\mathcal{PT}$ symmetry ($\mathcal{P}$: spatial inversion and $\mathcal{T}$: time-reversal symmetry), the Fermi surfaces are spin split and therefore not spin degenerate.
In Fig.~3(b), the experimental frequencies are compared with those expected from these Fermi surfaces.
The $\alpha$ and $\zeta$ branches are most naturally assigned to band 60, the $\beta$ and $\gamma$ branches to band 61, and the $\varepsilon$ branch to band 59, while the $\delta$ branch likely contains contributions from both bands 63 and 59. 
The $\beta$ branch may alternatively originate from band 60.
The frequencies lower than those of the $\alpha$ and $\beta$ branches are likely associated with band 60.
The frequencies around $F = 800$--900~T observed near $\phi = 0^\circ$ are attributed to the second harmonics of the $\gamma$, $\delta$, and $\varepsilon$ branches.
In addition, the frequencies around $F = 500$--600~T observed over a wide angular range may be associated with band 61 and/or band 62, although a definitive assignment cannot be made at present.

The effective masses of the major frequency branches for $B \parallel a$ were determined from the temperature dependence of the oscillation amplitudes (see the Supplemental Material for details).
Table~I compares the experimental effective masses with the band masses obtained from the band-structure calculations.
The effective masses of the $\gamma$, $\delta$, $\varepsilon$, and $\zeta$ frequencies are around $1.4\,m_e$, whereas the band masses of their candidate orbits are around $0.7\,m_e$, except for the light orbit at the X point of band 63.
Therefore, the effective masses are enhanced by approximately a factor of two relative to the band masses owing to electron--phonon coupling and electron--electron correlations.
It is useful to compare this mass enhancement with those reported for other magnetic Mn compounds.
The mass enhancement factors have been reported to be 2.8 for MnP \cite{Takase79JPSJ, Kawakatsu19JPSJ}, 5.0 for MnSi \cite{Wilde21Nature}, 5 for Mn$_3$Sn \cite{Kuroda17NatMater}, and 1.2 for Mn$_3$Ge \cite{Changdar24JPCM}.
The mass enhancement of approximately 2 observed in NbMnP therefore falls within a reasonable range.

Taken together, the low-field SdH oscillations in NbMnP are reasonably well described by the band-structure calculations based on the $B_{3g}+B_{2u}$ magnetic structure.

\begingroup
\squeezetable
\begin{table}
\caption{Comparison of the experimental SdH frequencies and effective masses with the calculated frequencies and band masses in NbMnP for $B \parallel a$.
The experimental values are averages obtained from three samples, A, O, and R.
The coordinates $k_x$, $k_y$, and $k_z$ denote the centers of the corresponding orbits and are expressed in units where $2\pi/a$, $2\pi/b$, and $2\pi/c$ are normalized to 100, respectively.
}
\begin{ruledtabular}
\begin{tabular}{ccccccccc}
\multicolumn{3}{c}{Experiment} & \multicolumn{6}{c}{Calculation} \\
\cline{1-3} \cline{4-9}
Branch & \multicolumn{1}{c}{$F$ (T)} & \multicolumn{1}{c}{$m^*/m_e$} & \multicolumn{1}{c}{Band} & \multicolumn{1}{c}{$F$ (T)} & \multicolumn{1}{c}{$m_{band}/m_e$} & \multicolumn{1}{c}{$k_x$} & \multicolumn{1}{c}{$k_y$} & \multicolumn{1}{c}{$k_z$}  \\
\hline
$\alpha$ & 92 & 0.69(5) & 60 & 126 & 0.79 & 16 & 1 & 11 \\
$\gamma$ & 402 & 1.36(5) & 61 & 364 & 0.72 & 21 & 0 & 50 \\
 &  &  & 59 & 414 & 0.83 & 0 & 32 & 0 \\
$\delta$ & 437 & 1.51(6) & 63 & 440 & 0.42 & 50 & 0 & 0 \\
$\epsilon$ & 499 & 1.28(6) & 59 & 530 & 0.75 & 0 & -32 & 0 \\
$\zeta$ & 573 & 1.43(5) & 60 & 590 & 0.76 & 0 & -32 & 0 \\
&  &  & 60 & 611 & 0.75 & 0 & 31 & 0 \\
\end{tabular}
\end{ruledtabular}
\end{table}
\endgroup

\begin{figure*}
\includegraphics[width=17cm]{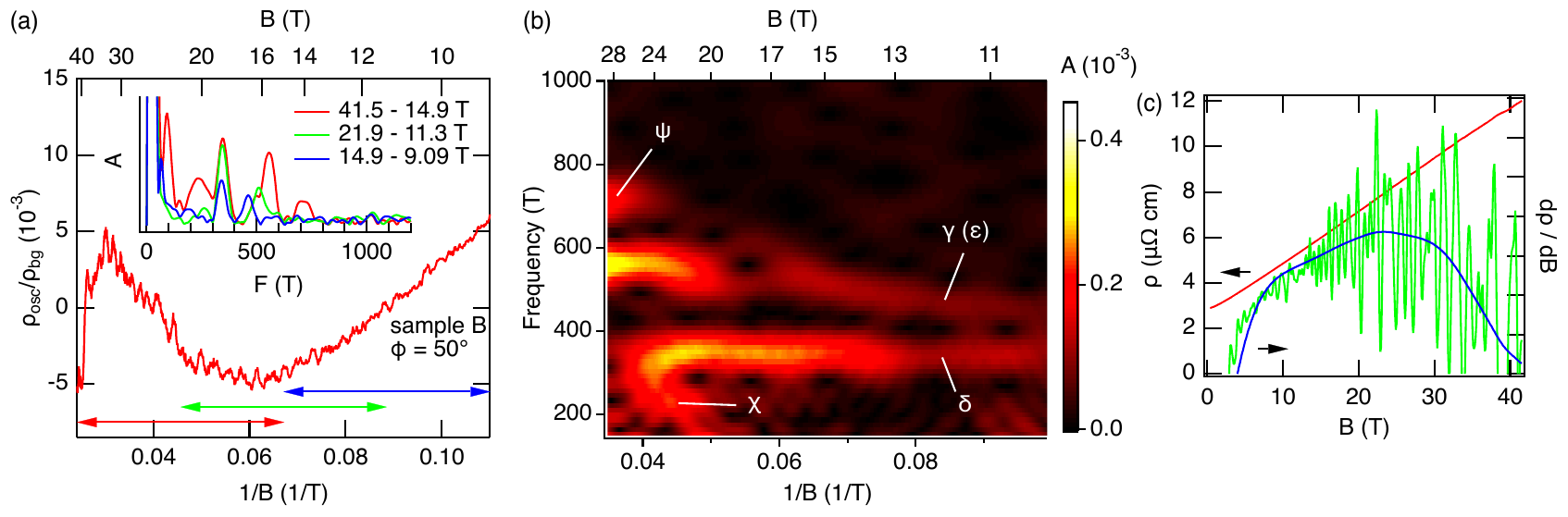}
\caption{\label{sample_B} 
(a) SdH oscillations at $\phi=50^\circ$ plotted as a function of $1/B$.
The inset shows the Fourier transform spectra obtained for the low-, intermediate-, and high-field regions.
(b) Spectrogram of the SdH oscillations shown in (a), obtained by a short-time Fourier transform.
The inverse-field window width was $\Delta(1/B)=0.0859$ T$^{-1}$.
Regions affected by end effects at both the low- and high-field sides have been removed from the spectrogram.
(c) Field dependence of the electrical resistivity at $\phi=50^\circ$, from which the SdH oscillations shown in (a) were extracted, together with its field derivative.
A smoothed version of the derivative is also shown to highlight the field dependence of the background component.
}
\end{figure*}

We now return to the issue of the field-induced evolution of the Fermi surface.

Figure~4(a) shows the SdH oscillations at $\phi=50^\circ$ plotted as a function of $1/B$.
The data were divided into three equal-width windows in $1/B$, corresponding to the low-, intermediate-, and high-field regions, and the resulting Fourier spectra are shown in the inset.
The spectra clearly depend on the field range used for the Fourier transform.

To examine the field evolution of the spectra in more detail, a short-time Fourier transform (STFT) was performed.
The resulting spectrogram is shown in Fig.~4(b).
The window width used for the transformation was $\Delta(1/B)=0.0859$ T$^{-1}$.
Regions affected by end effects at both the low- and high-field sides were excluded from the spectrogram.

At the lowest fields (right side of the figure), the $\delta$ frequency at $F \sim 340$~T and the $\gamma$ (or $\varepsilon$) frequency at $F \sim 450$~T are observed.
With increasing field, the $\delta$ frequency remains nearly unchanged, whereas the $\gamma$ (or $\varepsilon$) frequency gradually shifts to higher values.
The intensity of the $\gamma$ (or $\varepsilon$) branch is strongly suppressed near $1/B=0.055$ T$^{-1}$, although the origin of this suppression is unclear.
A new frequency $\chi$ lower than the $\delta$ branch appears near $1/B=0.05$ T$^{-1}$.
As the field increases (i.e., with decreasing $1/B$), it shifts rapidly to higher frequencies and merges with the $\delta$ branch near $1/B=0.04$ T$^{-1}$.
At still higher fields ($1/B \lesssim 0.04$ T$^{-1}$), the $\delta$ branch disappears, while a new frequency $\psi$ emerges near $F \sim 700$~T.

In itinerant magnetic systems, field-induced changes in magnetization and Fermi surface are inseparably linked.
For simplicity, let us consider an itinerant ferromagnet.
The magnetization is determined by the difference between the Fermi volumes of the majority- and minority-spin bands.
Therefore, an increase in magnetization under an applied magnetic field requires an expansion of the majority-spin Fermi surface and a contraction of the minority-spin Fermi surface.
In the present case, the noncollinear magnetic structure prevents such a simple majority/minority-spin description.
Nevertheless, it is reasonable to expect that the applied field induces local expansions and contractions of different parts of the Fermi surface, thereby changing the magnetization.
Within this scenario, the nearly field-independent $\delta$ frequency and the continuous shift of the $\gamma$ (or $\varepsilon$) frequency observed up to approximately $1/B \sim 0.06$ T$^{-1}$ can be naturally understood.

In contrast, the disappearance of the $\delta$ frequency and the emergence of the new frequency $\psi$ near $F \sim 700$~T around $1/B \sim 0.04$ T$^{-1}$ constitute clear evidence for a Lifshitz transition \cite{Lifshitz60JETP}, i.e., a change in the topology of the Fermi surface.

We therefore examine whether this transition leaves detectable signatures in the field dependence of the magnetoresistance.
Figure~4(c) shows the field dependence of the resistivity at $\phi=50^\circ$ from which the SdH oscillations in Fig. 4(a) were extracted, together with its field derivative.
A smoothed version of the derivative is also shown in order to suppress the SdH oscillations and reveal the field dependence of the smooth background component.
It can be seen that the background derivative changes from increasing to decreasing at approximately 23~T (close to $1/B \sim 0.04$ T$^{-1}$), indicating a change in the curvature of the background magnetoresistance from concave upward to concave downward.
This feature is likely associated with the Lifshitz transition.

Field-induced Lifshitz transitions have been reported in a variety of heavy-fermion systems \cite{Kozlova05PRL,Yelland11NatPhys,Aoki16PRL,Broad26NJP}.
Outside the heavy-fermion class, reported examples remain scarce and are mostly limited to low-carrier-density systems in connection with the quantum limit \cite{Galeski22NatCommun}.
The observation of such a transition in a system with a substantial carrier density, such as NbMnP, is therefore rare.
The spin-split electronic structure of NbMnP likely plays an essential role in enabling the field-induced Lifshitz transition.

In conclusion, we have observed Shubnikov--de Haas oscillations in high-quality single crystals of the antiferromagnet NbMnP exhibiting an anomalous Hall effect.
The low-field SdH frequencies are well accounted for by first-principles calculations based on the $B_{3g}+B_{2u}$ magnetic structure determined by neutron diffraction.
At higher fields, however, the SdH spectrum evolves markedly and reveals a field-induced Lifshitz transition.
This result demonstrates that field-induced modifications of the Fermi surface must be taken into account when discussing the high-field magnetotransport properties of itinerant antiferromagnets.

\begin{acknowledgments}
This work was supported by JSPS KAKENHI Grant Numbers JP22K03537, JP24KJ0227, JP26K17065, JP23K25827, JP24K00581, JP24K00588, JP25K00947, JP25K21684, JP25K00947, and JP23H04871.
MANA is supported by the World Premier International Research Center Initiative (WPI), MEXT, Japan.
The National High Magnetic Field Laboratory is supported by National Science Foundation through NSF/DMR-2128556 and the State of Florida.
We thank Troy Brumm for his valuable technical support at NHMFL.
\end{acknowledgments}


%

\end{document}